\documentclass[journal=jacsat,manuscript=article]{achemso}

\usepackage{chemformula} 
\usepackage[T1]{fontenc} 
\usepackage{amssymb}

\author{Srijani Mallik}
\affiliation{Unit\'e Mixte de Physique, CNRS, Thales, Universit\'e Paris-Saclay, Palaiseau, France}
\author{Gerbold C. Ménard}
\affiliation{Laboratoire de Physique et d’Etude des Matériaux, ESPCI Paris, Université PSL, CNRS, 75005, Paris, France}
\author{Guilhem Saiz}
\affiliation{Laboratoire de Physique et d’Etude des Matériaux, ESPCI Paris, Université PSL, CNRS, 75005, Paris, France}
\author{Alexandre Gloter}
\affiliation{Laboratoire de Physique des Solides, Université Paris-Saclay, CNRS UMR 8502, 91405 Orsay, France}
\author{Nicolas Bergeal}
\affiliation{Laboratoire de Physique et d’Etude des Matériaux, ESPCI Paris, Université PSL, CNRS, 75005, Paris, France}
\author{Marc Gabay}
\affiliation{Laboratoire de Physique des Solides, Université Paris-Saclay, CNRS UMR 8502, 91405 Orsay, France}
\author{Manuel Bibes}
\email{manuel.bibes@cnrs-thales.fr}
\affiliation{Unit\'e Mixte de Physique, CNRS, Thales, Universit\'e Paris-Saclay, Palaiseau, France}

\title{From low-field Sondheimer oscillations to high-field very large and linear magnetoresistance in a SrTiO\textsubscript{3}-based two-dimensional electron gas}

\begin{document}








\begin{abstract}
  Quantum materials harbor a cornucopia of exotic transport phenomena challenging our understanding of condensed matter. Among these, a giant, non-saturating linear magnetoresistance (MR) has been reported in various systems, from Weyl semi-metals to topological insulators. Its origin is often ascribed to unusual band structure effects but it may also be caused by extrinsic sample disorder. Here, we report a very large linear MR in a SrTiO$_3$ two-dimensional electron gas and, by combining transport measurements with electron spectro-microscopy, show that it is caused by nanoscale inhomogeneities that are self-organized during sample growth. Our data also reveal semi-classical Sondheimer oscillations arising from interferences between helicoidal electron trajectories, from which we determine the 2DEG thickness. Our results bring insight into the origin of linear MR in quantum materials, expand the range of functionalities of oxide 2DEGs and suggest exciting routes to explore the interaction of linear MR with features like Rashba spin-orbit coupling.
\end{abstract}




\vspace{3em}

Since their discovery in 2004 \cite{ohtomo_high-mobility_2004}, oxide two-dimensional electron gases (2DEGs) have been shown to possess a wide range of functionalities including superconductivity \cite{reyren_superconducting_2007} and Rashba spin-orbit coupling (SOC) \cite{caviglia_tunable_2010}. Consequently, these 2DEGs display a rich phenomenology of electron transport. They commonly show Lorentz magnetoresistance, weak localization and weak anti-localization due to the Rashba SOC \cite{caviglia_tunable_2010}). Further, samples with high mobilities exhibit Shubnikov-de Haas oscillations \cite{shalom_shubnikovhaas_2010,fete_large_2014} and quantum Hall effect  \cite{trier_quantization_2016}. A peculiar angular dependence of the resistance, including a unidirectional magnetoresistance \cite{choe_gate-tunable_2019,vaz_determining_2020} and high order oscillations \cite{joshua_gate-tunable_2013}, has also been reported \cite{trushin_anisotropic_2009,bovenzi_semiclassical_2017,fete_rashba_2012,vaz_determining_2020}.  

A few studies also mention a linear magnetoresistance (MR) for perpendicular magnetic field $B$ \cite{jin_large_2016,kormondy_large_2018} with, in some cases, a very large amplitude ($>$500\% at 9 T) \cite{jin_large_2016}. This is reminiscent of a giant, non-saturating linear MR reported in various systems including narrow band-gap semiconductors \cite{Hu2007}, graphene \cite{Wang2014,Kisslinger2015}, semimetals \cite{Liang2015, Huang2015, Narayanan2015}, and topological insulators \cite{Yan2013, Leng2020}. Two main scenarios have been proposed to explain this behavior. Abrikosov \cite{Abrikosov1998, Abrikosov2003} established that impurity scattering of Dirac fermions confined to the lowest Landau level, modelling quasi-gapless semi-conductors, results in a positive quantum MR varying linearly with $B$. Alternatively, Parish and Littlewood \cite{Parish2003, Parish2005} introduced a classical model describing the behavior of charges moving through an array of 4-terminal random resistors where they are subjected to a driving current along one direction and a magnetic field perpendicular to the array. These resistors represent regions where the mobility of the carriers fluctuate. The Lorentz force deflects the carrier trajectories, leading to Hall resistance in the local impedance matrix proportional to $B$. In the thermodynamic limit, when the width of the mobility distribution is less than the average, the network's impedance exhibits a positive MR that is linear in B and proportional to the average mobility above a cross-over field $B_C$ scaling with the inverse of the average mobility. Elaborating on this scenario, Kozlova \textit{et al.} \cite{Kozlova2012} considered the dynamics of high mobility two-dimensional classical electrons experiencing an in-plane longitudinal electric field and an out-of-plane magnetic field that are scattered off randomly placed, low mobility, large size islands. Multiple collisions on each island at the Hall velocity allow electrons to move around the obstacle and continue their course in the direction of the electric field. But this process causes an increase in the resistance that depends linearly on B above the cross-over field. It dominates the magneto-transport response of the system at high field.

In this Article, we report a very large linear MR in SrTiO$_3$ (STO) 2DEGs generated by sputtering an ultrathin film of Gd onto (001)-oriented STO crystals. We investigate the temperature and gate dependence of the slope of the MR and the cross-over field $B_C$ and find that they scale with the mobility as expected from the Parish-Littlewood model. Unlike in most previous reports of giant linear MR interpreted within this scenario, we provide a direct characterization of the nanoscale disorder responsible for the effect through various microscopy techniques and electron energy-loss spectroscopy. In addition, we report the observation of oscillations in the resistance and its field derivatives that are periodic in $B$. We argue that they correspond to Sondheimer oscillations \cite{sondheimer_influence_1950} and use them to extract a 2DEG thickness that agrees with the spectro-microscopy data.


We have deposited Gd films of various thicknesses by sputtering on STO (001) substrates. Prior to Gd deposition, the Ti valence of the STO substrates was studied \textit{in situ} using X-ray photoelectron spectroscopy (XPS). The inset of Figure 1(a) shows the spectrum collected for the  Ti 2$p$ core levels of the substrate. Adventitious carbon was used as a charge reference to obtain the Ti$^{4+}$ 2$p_{3⁄2}$ peak position $\sim$458.6 eV for the fitting. The energy difference between 2$p_{3⁄2}$ and 2$p_{1⁄2}$ was constrained to 5.7 eV. These values are consistent with values reported previously for STO \cite{sing_profiling_2009,vaz_tuning_2017,vaz_mapping_2019}. We observed the sole presence of Ti$^{4+}$ states indicative of stoichiometric STO, as expected. Further, Ti 2$p$ core level spectra were measured \textit{in situ} after depositing Gd. Figure 1(a) reveals additional peaks corresponding to Ti$^{3+}$ and Ti$^{2+}$ states reflecting the population of the Ti $t_{2g}$ levels and the formation of the 2DEG. The relative concentration of Ti$^{3+}$ and Ti$^{2+}$ increases upon increasing the Gd thickness t$_{Gd}$ (Figure 1(c)), consistent with the situation observed in Y/STO heterostructures \cite{vicente-arche_metalsrtio3_2021}. The study for Gd 3$d$ core levels (figure 1(b)) reveals that a 1.55 nm Gd film gets fully oxidized into Gd$_2$O$_3$. Therefore, it is clear that Gd reacts with the surface oxygen atoms of the STO substrate and gets fully oxidized by the creation of oxygen vacancies into STO. In a previous report \cite{vicente-arche_metalsrtio3_2021} we showed that the degree of oxidation of various metals grown onto STO depends on their oxygen formation enthalpy and work function \cite{posadas_scavenging_2017,vaz_tuning_2017}. Here, we see that, owing to its much lower work function and more negative oxygen formation enthalpy in comparison to other metals (e.g. Al, Ta, Y) Gd induces a higher concentration of oxygen vacancies in STO. The solid circles in figure 1(c) show the Ti concentration measured after exposing the sample to the air. Clearly, the amount of Ti$^{3+}$ and Ti$^{2+}$ decreases due to the reoxidation of the 2DEG, consistent with results from \cite{vicente-arche_metalsrtio3_2021}. Therefore, we deduced the critical thickness of Gd to avoid the reoxidation of the reduced Ti from the air and thus for all the samples studied t$_{Gd}>$4 nm.

\begin{figure}[h!]
  \includegraphics[width=\linewidth]{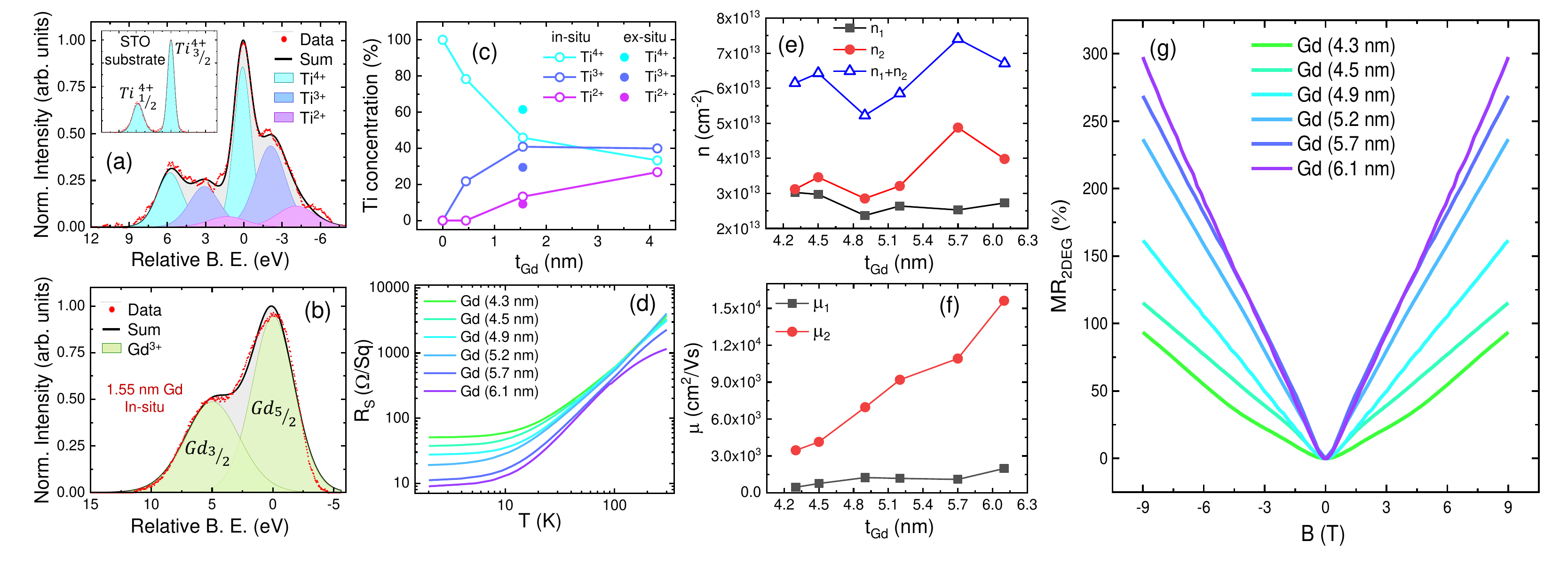}
  \caption{(a) X-ray photoelectron spectra (XPS) near the Ti 2$p$ edge for a bare SrTiO$_3$ single crystal substrate (inset) and after deposition of 1.55 nm of Gd. The fitted peaks for Ti\textsuperscript{4+}, Ti\textsuperscript{3+}, Ti\textsuperscript{2+}, and the sum fit envelope are shown in cyan, light blue, magenta colors and solid black line, respectively. Two Ti peaks of same color correspond to the 3/2 (right) and 1/2 (left) splitting of that specific Ti valence state. (b) XPS spectra near Gd 3$d$ edge after deposition of 1.55 nm of Gd. The fitted peaks with only Gd$^{3+}$ states suggest the oxidation of the whole Gd layer. (c) Concentration of different Ti reduced states as a function of deposited Gd thickness. The hollow and the solid circles represent the state of the sample before and after exposed to the air, respectively. (d) Temperature dependence of the sheet resistance of GdO\textsubscript{X}/STO samples of different thicknesses. (e) – (f) Carrier densities and the mobilities of the same samples extracted by fitting the Hall measurements and zero field sheet resistance at 2K with two band model. (g) Magnetoresistance measurements of 2D electron gas (2DEG) present in the GdO\textsubscript{x}/STO samples. The contribution of the 2DEG has been extracted by subtracting the metal contribution (if any) of Gd in the samples.}
  \label{fig:boat1}
\end{figure}

To confirm the generation of a 2DEG at the Gd/STO interface we measured the temperature dependence of the sheet resistance $R_S$ for samples having various Gd thicknesses (Figure 1(d)). A metallic behavior is observed for all samples. $R_S$ drops by almost two orders of magnitude from 300 K to 2 K, which validates the formation of 2DEG as inferred from XPS. At 2 K $R_S$ decreases monotonically upon increasing t$_{Gd}$. At 2 K the Hall traces for these samples were non-linear suggesting that transport proceeds in a multi-band regime. We extracted the carrier densities ($n_1$ and $n_2$) and mobilities ($\mu_1$ and $\mu_2$) of the 2DEGs by fitting the Hall data and the sheet resistance with a 2-band model \cite{kane_parallel_1985}, cf. Figure 1(e) and (f). While the carrier densities do not change much with t$_{Gd}$, the mobilities increase dramatically. Typical mobilities of metal/STO 2DEGs are within the range of 1000 cm$^2$/Vs \cite{vicente-arche_metalsrtio3_2021} but reach values as high as 15000 cm$^2$/Vs for t$_{Gd}$ = 6.1 nm. Assuming parabolic dispersion we extract for the 6.1 nm sample a mean free path $l_1$ = 0.13 $\mu$m  (resp. $l_2$ = 1.6 $\mu$m) for the low (resp. high) mobility carriers. Since $l_1$ and $l_2$ are quite large, we expect a semi-classical kinetic description of transport, allowing us to use the scaling established in Refs. \cite{Parish2003, Kozlova2012}. The MR curves at 2 K are shown in Figure 1(g). The large MR (in the 200 \% range) displays an unusual linear behavior and does not appear to saturate with $B$. All samples display a cross-over ($B_C$) between a low-field parabolic MR and a high-field linear MR (particularly clear for the 4.3 nm sample). Both the MR value and the dominance of the linear behavior increase with t$_{Gd}$. 

\begin{figure}[h!]
  \centering
  \includegraphics[width=0.8\linewidth]{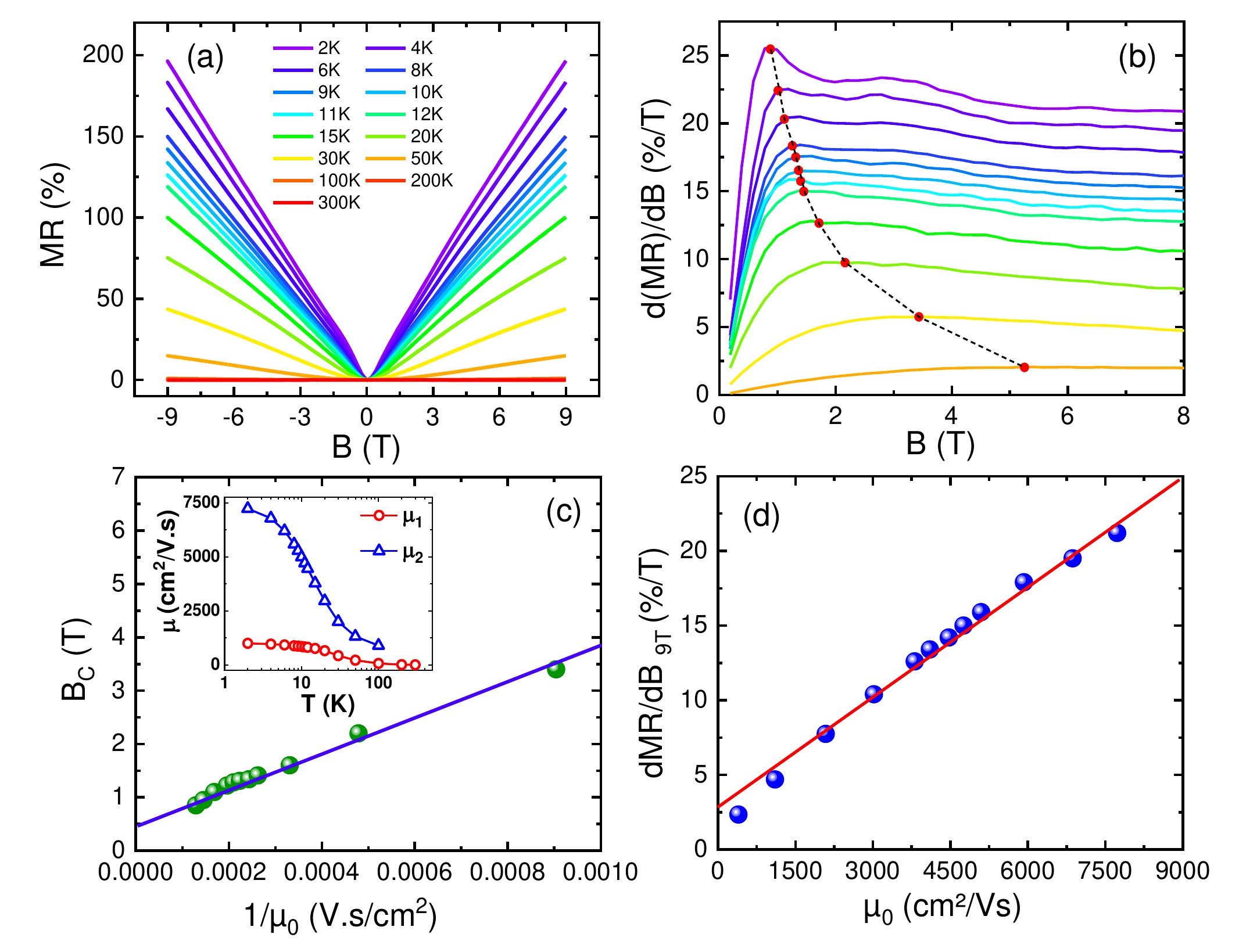}
  \caption{(a) Temperature dependent magnetoresistance curves for GdO\textsubscript{x} (6.1 nm)/STO sample. (b) First derivative of the magnetoresistance curves (shown in (a)) with respect to the magnetic field. The cross-over field ($B_C$) where the transition occurs from Lorentz MR to linear MR are pointed by the red solid circles on each curves. The color scheme for the temperatures is the same as in (a). (c) Cross-over field vs inverse of the zero field mobility measured at various temperatures. Inset: temperature dependent mobilities extracted from the fits of the Hall measurements using the two band model. (d) First derivative of MR at 9T as a function of zero field mobilities for different measurement temperatures. The blue and red straight lines connecting all the data points in (c) and (d), respectively suggest that scattering phenomena for all the temperatures is dominated by a single mechanism.}
  \label{fig:boat2}
\end{figure}

Figure 2(a) depicts the temperature dependent MR data for a 6.1 nm sample. The MR decreases upon increasing the temperature but its linear nature persists up to at least 50 K. This suggests that the MR has a classical origin as the effect is not observed at low temperature only. A cross-over between parabolic and linear behavior is again clearly visible. Figure 2(b) shows the first derivative of the temperature dependent MR curves where $B_C$ values are pointed by red solid circles. The low temperature data also reveal oscillations that we will discuss later. We have also measured the Hall effect at various temperatures and extracted the carrier densities and mobilities (cf. inset of Figure 2(c)). To test the applicability of the Parish-Littlewood model to our samples we plot the dependence of the cross-over field with the inverse of the zero field mobility ($\mu_0$) in Figure 2(c): as expected a linear dependence is obtained. Figure 2(d) shows that the slope of the linear MR scales with $\mu_0$, also consistent with Parish and Littlewood model. Interestingly, these data provide an estimate of the fraction $f$ of low-mobility islands in the sample. The slope of the MR in the linear regime is given by $\frac{dMR}{dB}\sim \mu^*B$ \cite{Kozlova2012}, where the effective mobility $\mu^*=\mu_0(\frac{f}{2(1-f)})$. From Fig. 1(g), we get $\frac{dMR}{dB}\sim 0.22$ /T  and from Fig. 2(b)-(c) we determine $\mu_0\sim 0.62$ m$^2$/Vs at 2K. Accordingly, we find $f\sim 0.4$. 

\begin{figure}[h!]
  \centering
  \includegraphics[width=0.8\linewidth]{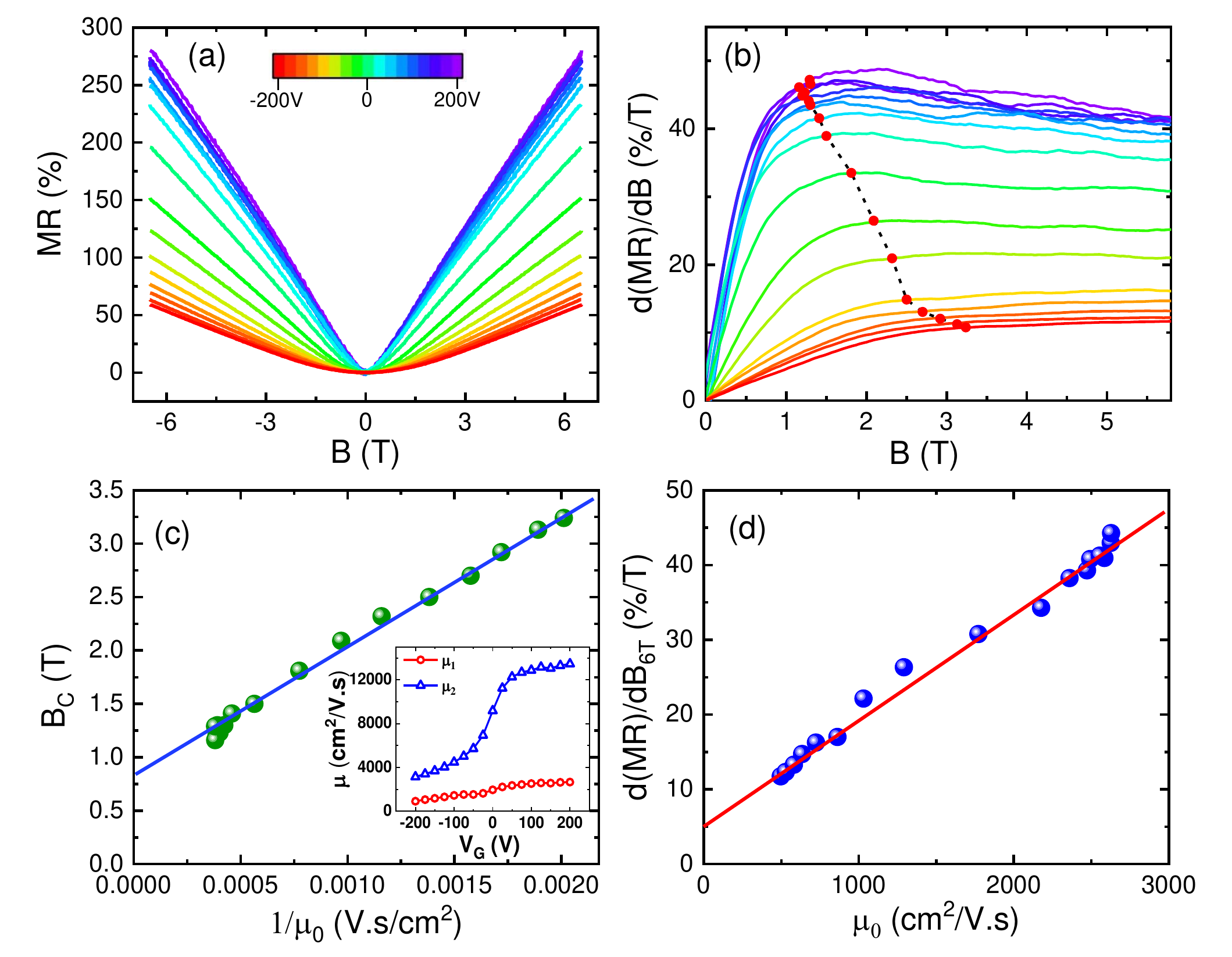}
  \caption{(a) Gate voltage dependent magnetoresistance curves for GdO\textsubscript{x} (6.1 nm)/STO sample at $T$ = 2 K. (b) First derivative of the magnetoresistance curves (shown in (a)) with respect to the magnetic field. The cross-over field ($B_C$) corresponds to the red solid circle on each curve. The color scheme for the gate voltages is the same as in (a). (c) Cross-over field vs inverse of the zero field mobility for different gate voltages. Inset: gate voltage dependent mobilities extracted from the fittings of the hall measurements using the two band model. (d) First derivative of MR at 6 T as a function of the zero field mobilities for different gate voltages. The blue and red straight lines connecting all the data points in (c) and (d), respectively suggest that scattering phenomena for all the gate voltages is dominated by a single mechanism.}
  \label{fig:boat3}
\end{figure}

Figure 3(a) shows the back gate voltage (V$_G$) dependence of magnetoresistance at $T$ = 2 K. For V$_G >$ -75V the linear MR dominates over the parabolic MR. However, for -75V $>$V$_G >$ -200V the parabolic MR becomes more prominent. As for the data of Figure 2, the cross-over field values were calculated from the first derivative of the MR curves (Figure 3(b)). The carrier densities and the mobilities were also calculated using a similar two band model. The inset of Figure 3(c) shows the gate dependence of the mobilities, revealing a sharp transition/increase around V$_G$ = -75V. This may indicate a Lifshitz transition to another band, correlated to the transition from the parabolic to linear MR. Consistent with the temperature dependent data, the cross-over field scales linearly with the inverse of the zero field mobility (Figure 3(c)). Further, the value of the first derivative of the MR at 6 T scales linearly with the zero field mobilities for the respective gate voltages (Figure 3(d)). This confirms the classical origin of the linear MR behavior in our system and supports our interpretation in the framework of the Parish-Littlewood model.

\begin{figure}[h!]
  \centering
	\includegraphics[width=0.6\linewidth]{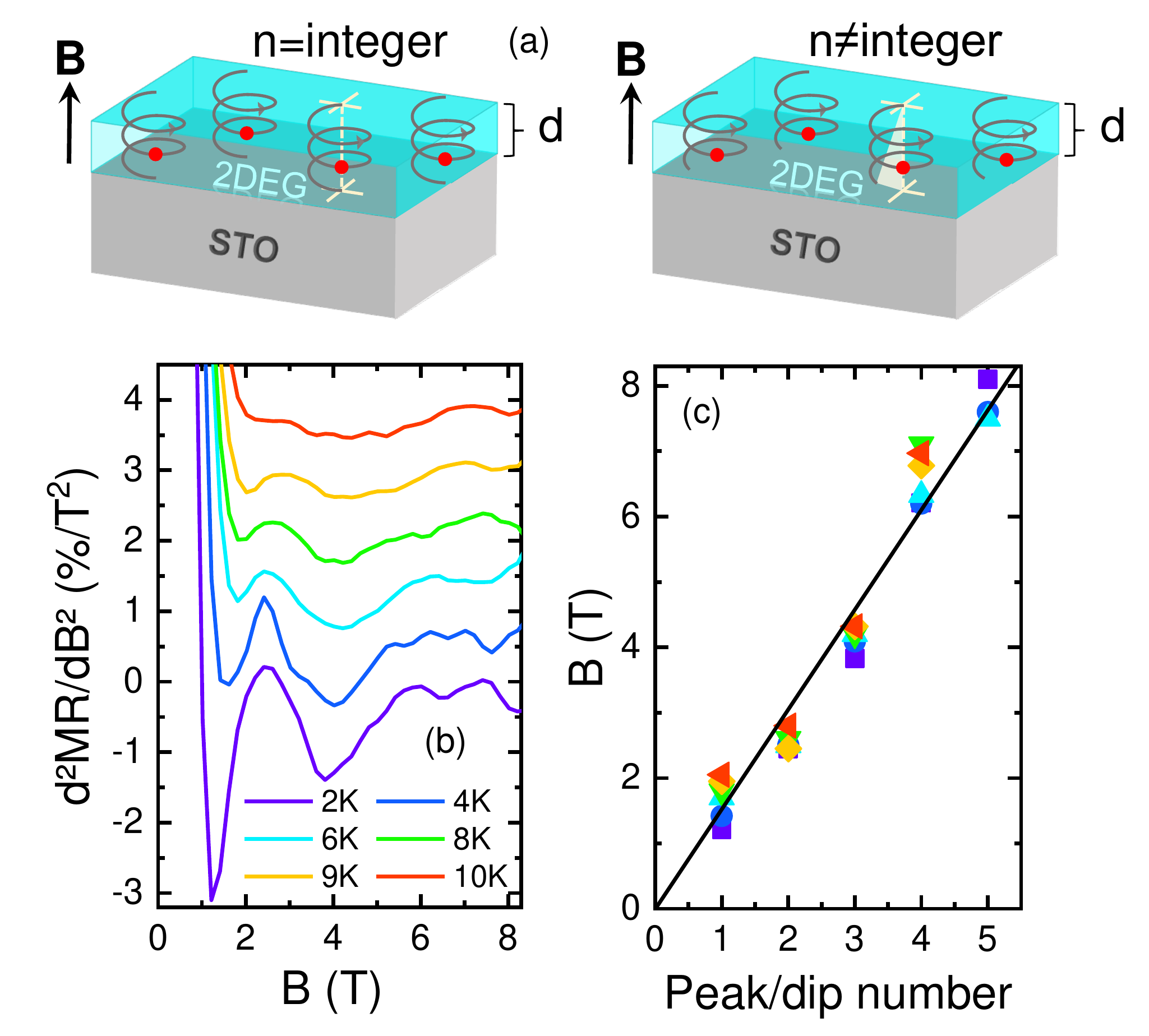}
	\caption{(a) Schematic representation of the mechanism of Sondheimer oscillations in STO 2DEGs. $d$ and $B$ represent the thickness of the 2DEG and the applied magnetic field, respectively. The red solid circles are the electrons and their helical motion in the magnetic field in z-direction are represented as grey lines. If the number of revolutions ($n$) needed for an electron to go from the top surface to the bottom is an integer then no net motion of the electron should be seen in the x-y plane (left). On contrary, if $n$ becomes non-integer then a net motion of the electron is seen in the x-y plane (right). (b) Sondheimer oscillations from second derivative of the magnetoresistance curves for $T$ = 2 to 10 K. (c) Peak and dip number with respect to their positions i.e. corresponding magnetic field values.}
	\label{fig:boat4}
\end{figure}

We now discuss the oscillations observed in the derivative of the MR displayed in Fig. 2b. When a magnetic field is applied perpendicular to the applied current, electrons perform cyclotron motion, leading to the quantization of the Fermi surface through the appearance of Landau levels. In this regime the resistance displays maxima and minima as the magnetic field is varied, called Shubnikov-de Haas (SdH) oscillations that appear at high field with a period scaling with $1/B$. In Fig. 4(b) we plot the second derivative of the MR in which the oscillations are more visible. They are present at low field and their period roughly scales with $B$ (cf. Fig. 4(c)). Thus, they cannot be ascribed to SdH oscillations. When undergoing cyclotron motion in the plane perpendicular to magnetic field, electrons may also acquire a net velocity in the magnetic field direction. The electrons then follow helical trajectories, which, as described by Sondheimer \cite{sondheimer_influence_1950}, may also produce resistance oscillations. Figure 4(a) depicts the schematic representation of Sondheimer oscillations in a conductive slab. For a given slab thickness and perpendicular electron velocity, the radius of the helical path and the number of revolutions ($n$) to travel from one surface to the other is determined by the applied magnetic field. If $n$ is an integer (top left figure 4(a)) no relative motion of the electron is observed in the plane. However, when $n$ is non-integer (top right figure 4(a)) a net in-plane motion (displacement with respect to the initial position in-plane) occurs. This results in an oscillatory MR behavior. Interestingly, the thickness of the conductive slab – here the 2DEG – can be determined from the period of the oscillations. To do that, we consider the time needed for the carriers to undergo one cyclotron revolution that is $t_C = 2\pi/\omega_C = 2\pi m/eB$ (where $\omega_C$, $e$ and $m$ are the cyclotron frequency, electron charge and electron mass, respectively) and the time needed to cross the 2DEG vertically is $t = d/v_z$ (where, $d$ is the thickness of the 2DEG and $v_z$ is the velocity of electrons along the z direction). A Sondheimer oscillation occurs when $t = n \times t_C$ and thus $v_z = e \Delta B d/2\pi m$. Now, considering a cylindrical Fermi surface, $v_z = \hslash k_F^z/ m_z = \hslash /m_z d$ where $m_z$ is the effective mass of electrons along $z$. By equating $v_z$ from both equations, $d = \sqrt{hm/\Delta B e m_z}$ where $\Delta B$ = 3T from Figure 4(b) and $m_z/m$ = 11 from both the out-of-plane and in-plane (not shown) magnetotransport data. Therefore, we determine the thickness of the 2DEG to be $\sim$10 nm.   

We finally return to the origin of the linear MR and to our interpretation within the Parish-Littlewood model assuming inhomogeneous transport. From our extraction of the mobilities and mean free paths, we estimate the average distance between the inhomogeneities to be of order $l_1\approx$130 nm. Let us now compare this number with values deduced from structural characterization.

As visible from the atomic force microscopy (AFM) image shown in Fig. 5a, Gd/STO samples have a peculiar surface morphology. While LAO/STO or AlO\textsubscript{x}/STO 2DEG samples are usually atomically smooth, here AFM reveals the presence of mesoscopic outgrowths with a height of around 30 nm and a diameter of 50-100 nm. We note that their average distance is comparable to the estimate mentioned above. Yet, the height and density of these outgrowths is particularly surprising given the thickness of the deposited Gd, here only 6.1 nm. 

To gain insight into their nature, we have performed transmission electron microscopy. Fig. 5b shows a cross section through three outgrowths. They appear to be hollow, having the same morphology as blisters, explaining the apparent inconsistency between their large height and the low deposited Gd thickness. Elemental analysis reveals the presence of a $\sim$2.5 nm thick amorphous layer of GdO\textsubscript{x} with some interdiffused Ti over the whole STO substrate and of a $\sim$6 nm layer of crystalline GdO\textsubscript{x} forming blisters. Below the blisters, the amorphous layer is quite homogeneous and the STO/Ti:GdO\textsubscript{x} interface is well defined. Beside the blisters, the amorphous layer is more discontinuous and the STO/Ti:GdO\textsubscript{x} interface is rougher. A possible mechanism leading to the formation of these mesotructures is the volume expansion and the stress modification of the Gd layer during its oxidation by the STO substrate \cite{keller_nano-scale_2019,tranvouez_effect_2020}. 

\begin{figure}[h!]
	\includegraphics[width=\linewidth]{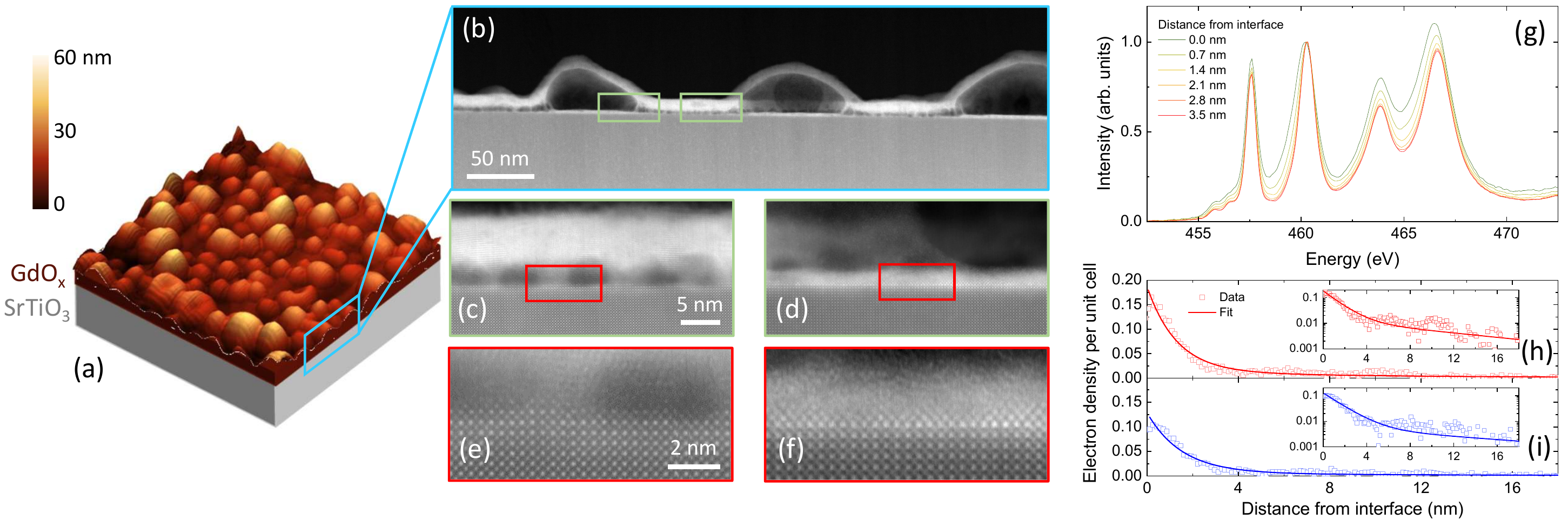}
	\caption{(a) 3D view of the topography of GdO\textsubscript{x}/STO sample by atomic force microscopy exhibits bubbles of GdO\textsubscript{x} due to oxidation. (b) low magnification high-angle annular dark field (HAADF) scanning transmission electron microscopy (STEM) image at the cross section (marked by the blue box) of the GdO\textsubscript{x} blisters, (c)-(d) HAADF images of the interface beside and below the bubble, respectively as marked by the green boxes, (e)-(f) inset zoom on the last STO plane for beside and below the bubble, respectively as marked by the red boxes, (g) electron energy loss spectroscopy (EELS) spectra showing typical Ti-$L$ edges evolution over the 4 nm on top of the STO, (h)-(i) charge profile as a function of the distance from the interface inside STO for below and beside the bubble, respectively. It shows a slightly sharper feature below the bubble in accordance with the sharper interface shown in (f). The insets are the zoom on the tail of the corresponding depth profiles.}
	\label{fig:boat5}
\end{figure}

The chemical analysis reveals differences between the regions located below or beside the blisters, with sharper Gd and Ti profiles below the blisters. The EELS fine structure of the Ti $L_{3, 2}$ edges (Fig. 5(g)) evidences that the Ti valence is different from pure 4+, consistent with XPS. Fits of the EELS spectra with reference components allow extracting the fraction of Ti$^{3+}$ (i.e. the electron density). The corresponding profiles below and beside the blisters are plotted in Fig. 5(h) and (i), respectively. One first notices that the electron density is higher below the blisters than beside them, which suggests that below the blisters the mobility is lower \cite{trier_electron_2018}, thus clarifying the origin of the inhomogeneous transport. Analysing the height distribution of the AFM image, we estimate that the outgrowths fill $\sim36 \%$ of the sample surface, which matches well the fraction of low-mobility regions extracted from the MR data (40$\%$). 

Both electron density profiles can be well fitted with the sum of two decaying exponentials, with characteristic decay lengths of 1.3 nm and 10 nm below the blisters and 1.5 nm and 14.3 nm beside the blisters. The shortest decay lengths match well the estimates of the 2DEG thickness for e.g. LAO/STO \cite{sing_profiling_2009}, where 2DEG formation is believed to mostly arise from interfacial charge transfer, that may also be at play here. The longest decay lengths are indicators of the total 2DEG thickness and are consistent with the generation of carriers through the formation of oxygen vacancies in the STO, created by the deposition of Gd. Remarkably, these decay lengths are in good agreement with our estimate of the 2DEG thickness from the Sondheimer oscillations ($\sim$ 10 nm).


In summary, we have observed a very large linear magnetoresistance in STO 2DEGs induced by the formation of 50-100 nm wide blisters during the deposition of a nanometric Gd film. This peculiar magnetotransport response could be investigated in more detail and more systematically in e.g. high-quality, homogeneous STO or KTaO$_3$ 2DEG samples patterned at the nanoscale to induce inhomogeneities with controlled dimensions and densities. In addition, we have identified the presence of Sondheimer oscillations, for the first time in oxide 2DEGs, and used their period to extract an accurate estimate of the 2DEG thickness. Sondheimer oscillations thus emerge as a new tool to investigate the physics of such systems, as has recently been done for other quantum materials \cite{van_delft_sondheimer_2020}.

\section{Materials and Methods}

\textbf{Sample preparation and XPS.} Gd thin films were deposited at room temperature by dc magnetron sputtering on as received STO (001) substrates from CrysTec. During Gd deposition, the Ar partial pressure and the dc power were kept fixed at $5.2\times10^{-4}$ mbar and 11 W, respectively. The deposition rate was $\sim$ 1 \AA/s. XPS was performed using a non-monochromatized Mg K$_\alpha$ source at 20 mA and 15 kV. Spectral analysis was carried out using CasaXPS.

\noindent \textbf{Magnetotransport.} Magnetotransport measurements were performed using a PPMS Dynacool system. The samples were bonded with Al wire in van der Pauw and Hall geometry for the magnetoresistance and Hall measurements, respectively. At the lowest temperature, the gate voltage was ﬁrst increased to its maximum value (+200 V) to suppress hysteresis and ensure the full reversibility of the measurements in the gate range  [-200 V, +200 V] \cite{singh_effect_2017}. The electron density and mobility of the two populations of electron were extracted by combining Hall effect and gate capacitance measurements as a function of gate voltage \cite{biscaras_limit_2015}.

\noindent \textbf{Structural analysis.} Surface topography was measured using atomic force microscopy. The average size of the Gd blisters and the area covered by these blisters were analyzed using Gwyddion. STEM was done using (monochromated) Cs-corrected Nion microscopes. The operating voltage was set to 100 keV in order to limit electron beam damage. EELS was collected with EMCCD detector and MerlinEM direct electron detector respectively mounted on a 3 and a 4/5 orders corrected EELS spectrometer. The EELS chemical analysis was done by analysing the Gd-$N$, Ti-$L$ and Gd-$M$ edges. The EELS valence quantification was done by fitting the Ti-$L$ edges with reference spectra (SrTiO$_3$ substrate at ca. 30 nm from the interface and a reference DyTiO$_3$ \cite{aeschlimann_living-dead_2018}).

\begin{acknowledgement}

The authors thank A. Vecchiola for collecting the AFM image and X. Li for helping with the STEM-EELS experiments. This project received funding from the ERC Advanced grant “FRESCO” $\#$833973 and from the H2020-EU “ESTEEM3” 823717.

\end{acknowledgement}



\providecommand{\latin}[1]{#1}
\makeatletter
\providecommand{\doi}
  {\begingroup\let\do\@makeother\dospecials
  \catcode`\{=1 \catcode`\}=2 \doi@aux}
\providecommand{\doi@aux}[1]{\endgroup\texttt{#1}}
\makeatother
\providecommand*\mcitethebibliography{\thebibliography}
\csname @ifundefined\endcsname{endmcitethebibliography}
  {\let\endmcitethebibliography\endthebibliography}{}

\end{document}